# Towards Participatory Design of Multi-agent

# Approach to Transport Demands

Yee Ming Chen[1] , Bo-Yuan Wang
Department of Industrial Engineering and Management
Yuan Ze University
135 Yuan-Tung Rd., Chung-Li, Tao-Yuan.
Taiwan, ROC.
[1]E-MAIL: chenyeeming@saturn.yzu.edu.tw

## Abstract

The design of multi-agent based simulations (MABS) is up to now mainly done in laboratories and based on designers' understanding of the activities to be simulated. Domain experts have little chance to directly validate agent behaviors. To fill this gap, we are investigating participatory methods of design, which allow users to participate in the design the pickup and delivery problem (PDP) in the taxi planning problem. In this paper, we present a participatory process for designing new socio-technical architectures to afford the taxi dispatch for this transportation system. The proposed dispatch architecture attempts to increase passenger satisfaction more globally, by concurrently dispatching multiple taxis to the same number of passengers in the same geographical region, and vis-'a-vis human driver and dispatcher satisfaction.

## 1. Introduction

Transportation systems are among the most ubiquitous and complex large-scale systems found in modern society. In many urban cities, public transportation systems are broadly accepted as the preferred transportation alternative for commuting to work, performing errands, or traveling to social events. Taxis are a convenient means of public transport in many countries. In providing quality customer service, fast and efficient fleet dispatching is essential [1,2]. The major focus of taxi dispatch systems has been primarily on reaching individual passengers in the shortest time possible to enhance passenger satisfaction. In the taxi system, most fares start with a phone call. A customer calls the taxi central and provides his or her location and desired destination. The taxi central assigns the delivery task to a taxi driver and drives his taxi to the customer. Once the customer gets aboard the taximeter starts running until the destination is reached. Abstracting from this normal situation we touch the topic of the Pickup and Delivery Problem (PDP) [3,4,5]. Various researches have been done in this field to optimize the performance or profit of the taxi companies. This means reaching the passengers via the shortest real-time paths possible. However, merely increasing individual passenger satisfaction, as is the current practice, is a local endeavor, in that it entails assigning the nearest taxi to a customer prioritized in a first come first serve queue, without considering the effects of the assignment on other awaiting customers in the request queue. To improve taxi fleet service performance, ideally, we should simultaneously and optimally assign taxis to service all customer bookings that are made within the time window. This is a challenging problem confronting current taxi dispatch systems.

The classic way to solve PDP's is to use heuristic algorithms [6]. This is a centralized approach in which a algorithms calculates the best routes for all the vehicles. It results in a fully optimized solution for delivering the goods in the minimum amount of time and costs with the resources available. The disadvantage is that you need to know all pickup and delivery points in advance. To deal with dynamic environment new approaches were introduced. One of them is the multi-agent approach. In multi-agent systems several agents work together to find the best solution for a problem [7].

In this paper, we focus on the multi-agent based simulations used for a class of problems that treats the real-time collaboration of human beings having very different perceptions and actions. We are implementing an agent-based simulator for this collaborative work. Each actor is modeled by an agent which is itself replaceable by a human actor in the simulations. We model the interaction between the multi- agent by using AnyLogic [8].

## 2. Taxi dispatch architecture

In the current centralized taxi dispatch architecture in use by a taxi operator in most of urban cities, incoming taxi service requests are queued on a first come first serve basis at the control center. For each passenger (mobile phone booking) request, available taxis in the vicinity of the passenger location are considered, and a taxi among them is dispatched to service it, upon the human driver accepting the booking job (Figure 1.). An efficient way is to assign a nearby taxi that can traverse the shortest-time path to the customer location, computed using real-time traffic information.

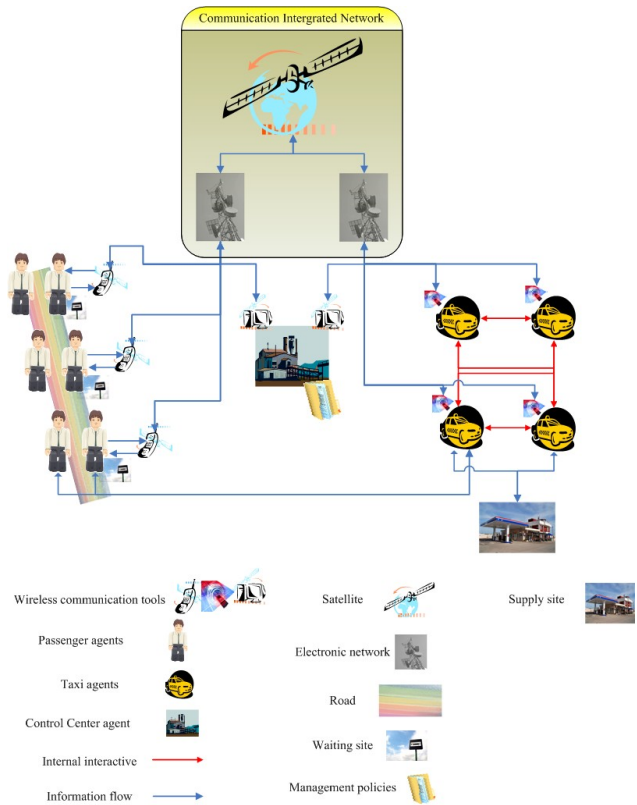

Figure 1. Taxi dispatch architecture

# 3. Simulation tool for participatory taxi dispatch system design

With the aim of integrating the experts in the design process, we have implemented an early version of the simulator to be designed, based on the multi-agent approach used for taxi dispatch system design. Towards this end, we propose using passenger, taxi, and control center agents capable of collaboration. And, to effectively utilize such software agents for taxi dispatch, we deploy them in a multi-agent architecture for PDP tasks, proposed as depicted in Figure 2.

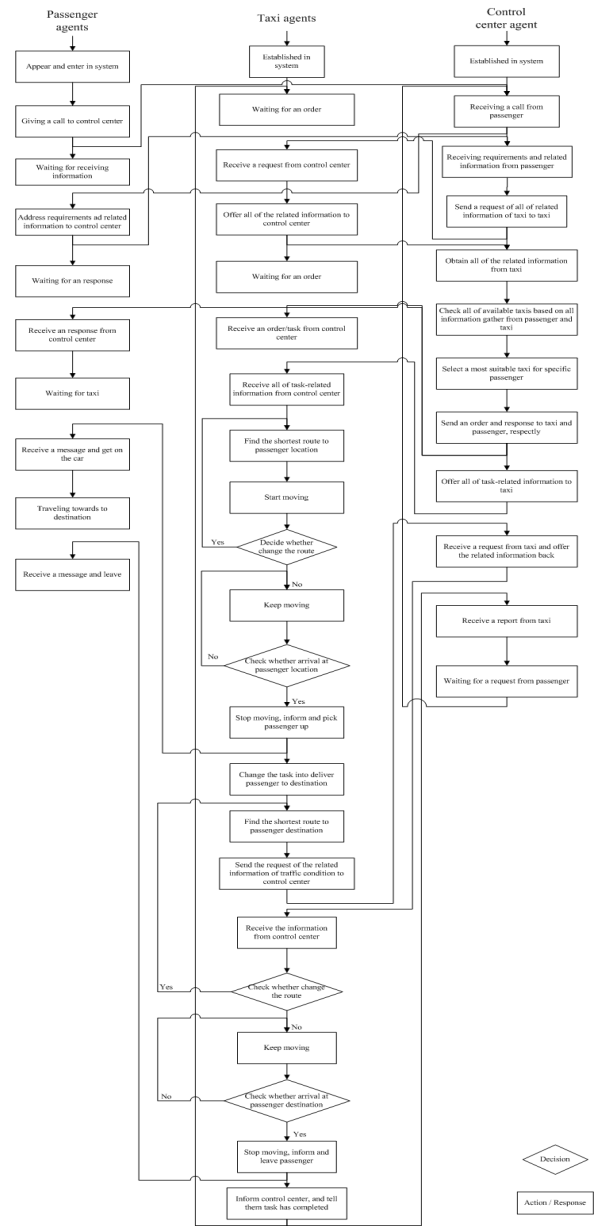

Figure 2. Proposed multi-agent based simulation for participatory taxi dispatch

## 3.1 Agents interface with users

An essential element for this type of "participatory taxi dispatch" is the existence of user interfaces. Their goal is not so much to help with the evaluation of the collaborative procedures but more to contribute so that the users can improve the agent behavior design [9,10]. They thus provide a perspective on the way in which the agents make their decisions. Each interface for a user is attached to an agent which can play alone the role of this user or can assist him. To achieve this, an attentive study of this agent capacity is needed for efficient human-agent (user-assistant) interactions.

A user plays the PDP task firstly to understand the agent decision-making mechanisms, secondly to validate this mechanism and finally to improve it. While executing the simulator, he observes the actions performed by other agents and his assistant. This human-agent relation leads to a kind of interactive learning [11,12]. To reach their learning application, a process consisting of three agents is envisaged (Figure 3):

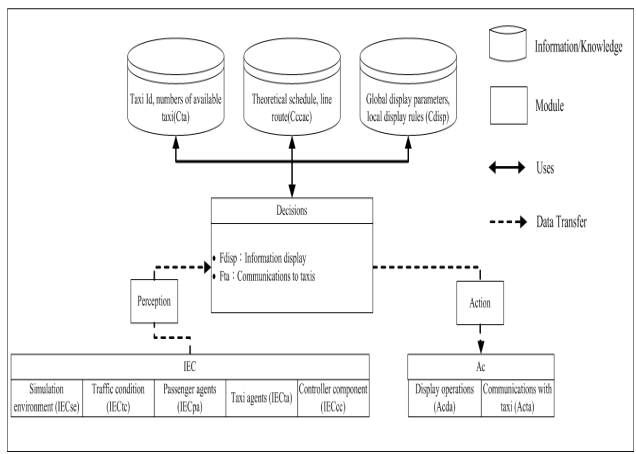

Figure 3(a) Definition of a Control Center agent

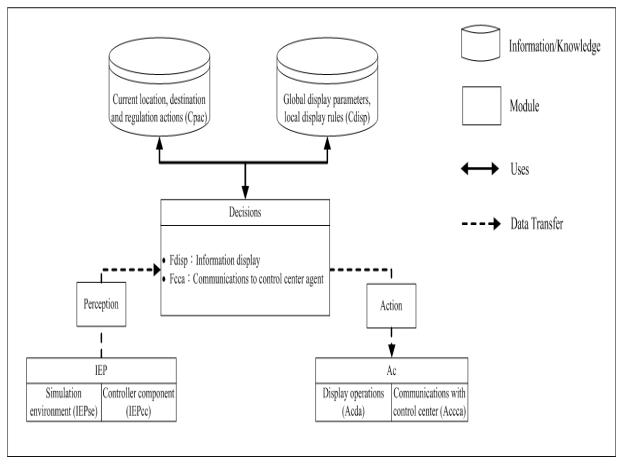

Figure 3(b) Definition of a passenger agent

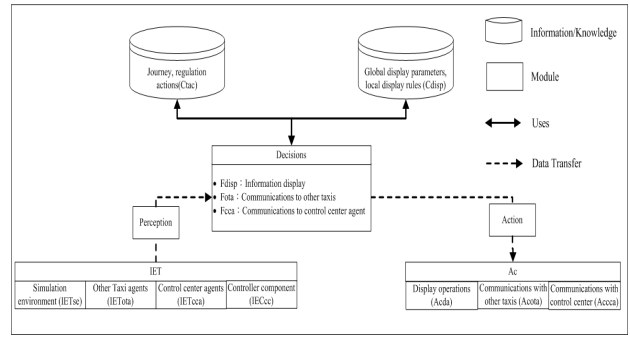

Figure 3(c) Definition of a taxi agent

We now summarize our major findings as the followings:

❑ The agent is designed to be able to build a log of validated/amended actions during interaction sessions with the user. The formed log must be structured in such a way the user can use it to correct agent behaviors.

❑ The action log is re-structured so that any agent can use it to learn its behaviors.

❑ The multi-agent are given the capacity to learn offline their behavior from the validated/amended actions conserved in the log.

❑ The learning techniques are applied in the on-line mode, *i.e.* during human-agent interactions.

All participating agents are in one of the designated areas of operation, and the taxi queues at the control center are updated accordingly. The communication protocol supporting the dispatch operations can then be prescribed as the figure 4.

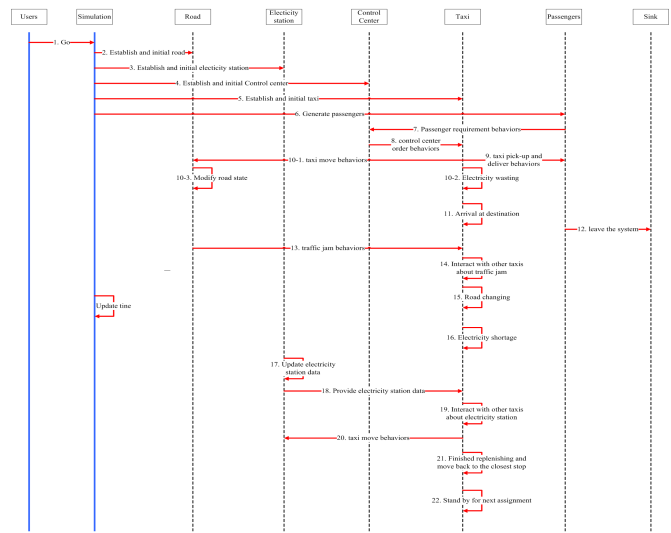

Figure 4  Participating agents interactions

## 4. Simulation and performance evaluation

To study the performance of the proposed multi-agent based simulation for participatory taxi dispatch, we conducted personal computer simulations on AnyLogic (Figure 5(a).) (http://kem.iem.yzu.edu.tw/), which simulating taxi operations in a selected urban road network of reasonable complexity. We focused on operational efficiency, in terms of passenger waiting time versus empty cruising time. In our simulations, passenger waiting time is measured from the moment a customer raises a request to the moment an assigned taxi arrives to pick up the passenger; empty taxi cruising time is measured from the moment it is available to the moment it accepts (or commits to service) a negotiated assignment.

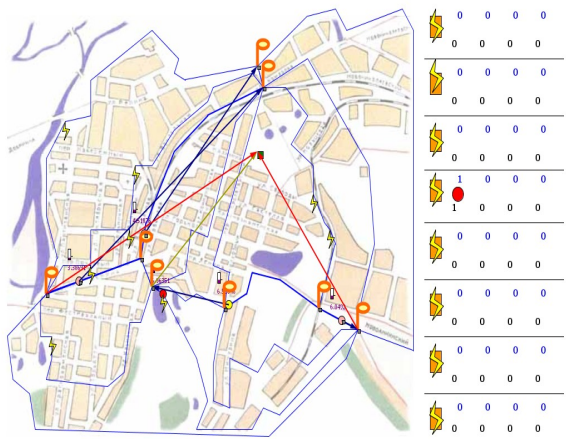

Figure 5(a).  The taxi dispatch within a road network in the proposed simulator

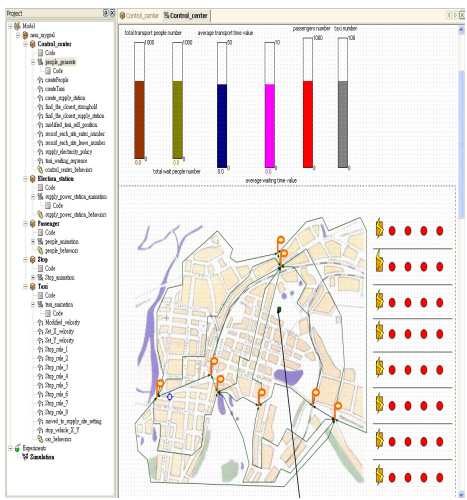

Figure 5(b).  The simulation of taxi PDP operation system

The simulation data (raw passenger waiting and empty cruising times)( (Figure 5(b).) gathered from the control center and taxi agents were saved into a text file for off line performance analysis.

The taxi company has to employ their taxis to satisfy the passengers around a city. In this city, taxi company provides PDP service mode. However, smaller taxi fleet scale will cause higher passenger idle time. Contrary, larger taxi fleet scale will cause higher taxi idle time. So what's the most suitable  taxi fleet scale is a very important issue for  taxi company. On the other hand, the path planning of taxi fleet also affects the passenger waiting time and taxi idle time. There are two existing paths planning: Shortest distance and least time. Therefore, what's the most suitable path planning for taxi fleet is also an important issue.

In order to deal with all of issues mentioned above, we have to construct the multi-agent based simulation for participatory taxi dispatch to simulate the operation of the participatory taxi system in this city. And our performance measures are passenger average waiting time, participatory taxi average idle time and participatory taxi average waiting time in queue. Then we will collect and analyze the data obtained from simulation. Finally, according to analysis results, we will make recommendations for the taxi company. The details of decision analysis for taxi PDP operation system simulation are described as follows.

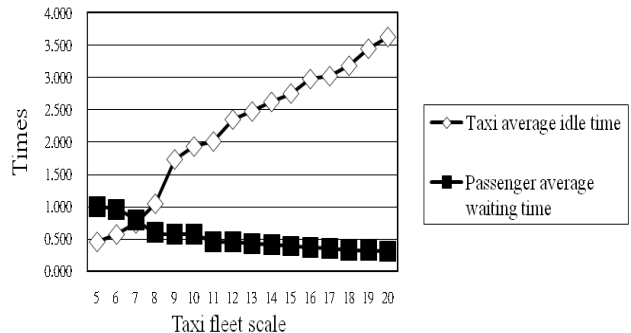

Figure 6  Taxi fleet scale

From figure 6, we can obviously realize that if the scale of taxi fleet is less than 7, the passenger average waiting time is higher than others. Contrary, if the scale of taxi fleet is more than 7, the taxi average idle time is too high. The most important thing is what the most feasible scale of taxi fleet is. We decide the scale of  taxi fleet is 11. The reason is that when taxi fleet scale is larger than or equals to 11, the passenger average waiting time begins converging.

Second, after deciding the scale of electric taxi fleet, the company wants to know which paths planning is better or more suitable for its' taxi fleet and taxi company operation. In order to find out which paths planning is better; we vary the participatory taxi number from 5 to 20 to observe the variances of passenger average waiting time and taxi average idle time.

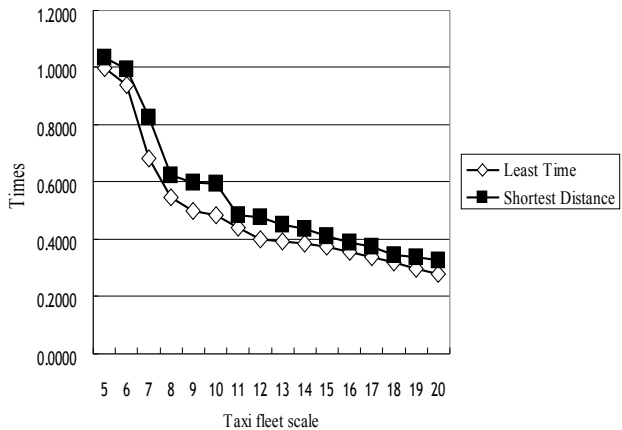

Figure 7   Passenger average waiting time

From figure 7, we can clearly know all of the passenger average waiting time is better under least time path planning. And there is a trend in figure 7. First, if the scale of taxi fleet increases, the passenger average waiting time will decrease. Contrary, if the scale of taxi fleet decreases, the passenger average waiting time will increase. Second, we find out that if the scale of taxi fleet is equal to or large than 11, the passenger average waiting time begins converging. In brief, the company adds more participatory taxi in the taxi fleet that only decreases small passenger average waiting time.

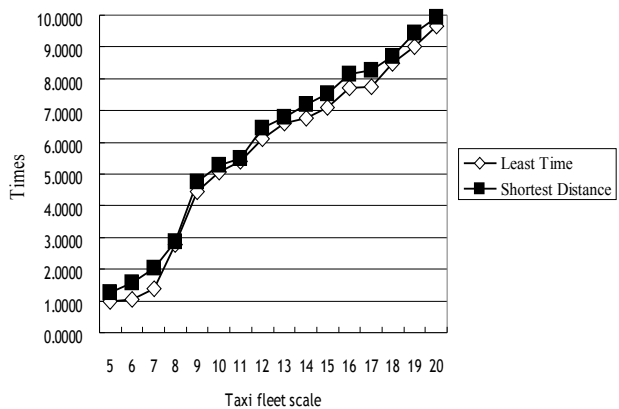

Figure 8   Taxi average idle time

In figure 8, we can clearly know most of the participatory taxi average idle time is worse under least time path planning. And there is a trend in figure 8. If the scale of taxi fleet increases, the participatory taxi average idle time also increases. Contrary, if the scale of taxi fleet decreases, the participatory taxi average idle time will decrease. So we can make a brief summary: The taxi company applies least time as path planning is more suitable for taxi fleet and passengers. Even though the taxi fleet scale increases causes the participatory taxi average idle time significantly increases, it also represents that the small scale of taxi fleet can achieve the lower passenger average waiting time.

# 5. Conclusions

The new idea of multi-agent based simulation for automating participatory taxi dispatch is introduced in this paper. The proposed participatory taxi dispatch architecture realizes the idea using collaborative assignment through a proposed multi-agent architecture. Using GIS simulations on an urban road network model, we evaluated the performance of the proposed dispatch system, and showed that, even on a basic infrastructure (Fig. 1), the distributed multi-agent system approach is promising in terms of improvements in operational efficiency over the existing centralized approach.

## Acknowledgements

This research work was sponsored by the National Science Council, R.O.C., under project number NSC97-2221-E-155-039.

**IJCSI**

**Yee Ming Chen** is a professor in the Department of Industrial Engineering and Management at Yuan Ze University, where he carries out basic and applied research in agent-based computing. His current research interests include soft computing, supply chain management, and system diagnosis/prognosis.

**Bo-Yuan Wang** is a graduated student in the Department of Industrial Engineering and Management at Yuan Ze University,, where he is studying basic and applied research in agent-based programming. His current research interests include web-based programming and system analysis/synthesis.